\documentclass[aps,superscriptaddress,nofootinbib,showpacs,eqsecnum,preprint,tightenlines]{revtex4}
\usepackage{hyperref}
\usepackage{epsfig,rotating}
\usepackage{amsmath,amssymb,graphicx}
\usepackage{dsfont}
\usepackage{bbm}
\numberwithin{equation}{section}

\newcommand{\be}{\begin{equation}}
\newcommand{\ee}{\end{equation}}
\newcommand{\bea}{\begin{eqnarray}}
\newcommand{\eea}{\end{eqnarray}}

\newcommand{\vk}{\vec{k}}

\begin{document}

\title{Coherence and decoherence in photon spin-qubit  entanglement.}

\author{Daniel Boyanovsky}
\email{boyan@pitt.edu} \affiliation{Department of Physics and
Astronomy, University of Pittsburgh, Pittsburgh, PA 15260}

\date{\today}

\begin{abstract}
We study the dynamics of spontaneous generation of coherence and  photon spin-qubit entanglement  in a $\Lambda$ system with non-degenerate lower levels. The cases of entanglement in frequency only and frequency and polarization are compared and  the reduced density matrix and   entanglement entropy are  analyzed. We explore in detail how   which-path information manifest when the energy difference between the qubit states is larger than the linewidth of the excited state suppresses coherence. A framework is provided to describe the   dynamics of spontaneous generation of coherence and (ideal) photodetection obtaining the post measurement qubit density matrix.     A  simple model of photodetection with a quantum eraser to suppress which-path information in the detection measurement is implemented. It is found that such quantum eraser purifies the   qubit density matrix after photodetection, our results are in agreement with those reported in recent experiments.

\end{abstract}


\maketitle

\section{Introduction}

Quantum entanglement has evolved from being a paradoxical aspect of quantum mechanics\cite{epr} to becoming a resource for quantum computing and quantum information\cite{bookqiqc,kimble,duan} with potential for technological breakthroughs in these areas\cite{expt1,horodecki,esentan}.  Several recent experiments demonstrated  photon entanglement with single atoms \cite{photatom1,photatom2,photatom3}, atomic ensembles\cite{ensemble}, long-distance entanglement between qubits\cite{entdistance,zoeller1,zoeller2}, and tunable ion-photon entanglement in optical cavities\cite{ions1,ions2}. Along with atom-photon entanglement\cite{photatom1,photatom2,photatom3,duan}, and entanglement in cavity quantum electrodynamicsl\cite{cavity} recent proposals suggested electron spin-photon entanglement in quantum dots as platforms for entanglement between distant spins\cite{diode}. Spin-photon entanglement  could be the pathway towards implementation of quantum networks among distant nodes\cite{kimble,zoeller1,zoeller2}.  Remarkable   experiments   demonstrated the realization of    entanglement between the polarization of a single optical photon and an electronic spin qubit in nitrogen vacancy (NV) centers in diamond\cite{dutt} and more recently the demonstration of   entanglement between a single electron spin and a photon in a quantum dot has been reported\cite{qdotyama,qdotgao,qdot}. A main paradigm in many of these experiments is that of spontaneous generation of coherence\cite{java,sham1,sham2} in a type-II or $\Lambda$ system, namely a situation in which spontaneous emission from a single
excited state  via a two-channel decay to   degenerate or non-degenerate lower levels results in coherence between these two states.  Spontaneous generation of electron spin coherence has also been observed from the radiative decay of charged excitons (trions) in quantum dots\cite{dotdutt}.

These experimental efforts are paving the way towards the implementation of atom-photon or spin-photon entanglement as   potential platforms for quantum information and quantum computing   protocols and networks\cite{imamoglu,kimble,duan}, motivating a theoretical effort seeking a deeper understanding of these processes\cite{shabaev,chen,sham1,mismatch}.

Although there have been some recent studies of the dynamics of spontaneously generated coherence\cite{sham1,sham2,mismatch} many important aspects merit further investigation.


Our main goal  in this article is to provide a more complete theoretical study of the experimental results reported in ref.\cite{dutt} but that apply more generally to current experiments on spin-qubit-photon entanglement\cite{qdotyama,qdotgao,qdot} from spontaneous generation of coherence as mentioned above.
 With this aim,   we focus on the following aspects: \textbf{1)}  to provide a  treatment of the dynamics of spontaneous generation of coherence, entanglement both in frequency and polarization and photodetection within a single framework  consistent with causality\cite{causality}, \textbf{2)} to study the entanglement entropy of reduced spin-qubit density matrices after tracing over the radiation degrees of freedom for photon-qubit entanglement both in frequency and polarization, of particular interest when spontaneous emission produces polarized photons which are measured by projection on differen polarization states \textbf{3)} to analyze in detail how which-
path information affects coherence, in particular within the setting of the experiment in ref.\cite{dutt}, predicting the time dependence of conditional probabilities when which path information is present.  \textbf{4)}To implement a model for a ``quantum eraser''\cite{eraser1,eraser2} within the framework of photodetection \emph{a l\'a} Glauber\cite{book1,book2,glauber} so as to erase which path information in the photodetection process.
An important result of this treatment is that ``quantum erasing'' ``which- path'' information leads to the purification of the qubit state confirming the experimental results of ref.\cite{dutt,qdotyama} and bolstering the arguments on ``quantum erasing'' in these references. We obtain a conditional probability in complete agreement with the experimental results of ref.\cite{dutt}. 

Our study differs from and complements recent theoretical treatments of spontaneous generation of coherence\cite{sham1,mismatch} in that we analyze both frequency and polarization entanglement, which-path decoherence, the spin-qubit entanglement entropy and incorporate a Glauber model of broadband photodetection[34,35,36] in a unified manner with the treatment of spontaneous emission. This treatment directly builds in causality in the spontaneous emission/photodetection process[31], leads to detailed understanding of how which path information affects coherence,  and allows  to model a quantum eraser[32,33]  consistently within the broadband photodetector model. This approach is different from that advocated in a recent article\cite{mismatch} where the photodetector is modeled with a collection of two-state atoms spread over some distance where the excited state features a short lifetime. Furthermore our study also differs from those of refs.\cite{sham1,mismatch} in that it shows  how the implementation of a ``quantum eraser'' leads to the purification of the qubit density matrix upon photodetection and yields a result for the conditional probability in complete  agreement with the experimental findings in ref.[19].

\section{Dynamics of entanglement via spontaneous decay.}
We   consider a $\Lambda$-system with one excited state $|A\rangle$ and two Zeeman split non-degenerate lower levels interacting with the electromagnetic field in the dipole and rotating wave approximations. The degenerate case can be obtained straightforwardly. We refer to the two-lower state levels $|\pm 1\rangle$  as a spin- qubit. The cases in which there is photon-qubit entanglement in frequency only and in frequency and polarization are studied separately and compared.

\subsection{Entanglement in frequency only:}\label{subsec:freqonly}
  We first consider the case when the dipole matrix elements are independent of the polarization of the photon and for simplicity we only consider one polarization to establish contact with the results of ref.\cite{sham1}.   This case leads to qubit-photon entanglement in frequency only, and generalization    to two polarizations is straightforward.  The total Hamiltonian for the three level system
is given by
\be H= H_A+H_R+H_{AR}\,, \label{totalH}\ee where
\be H_A = E_A |A\rangle \langle A|+ E_+ |+1\rangle \langle +1|+E_- |-1\rangle \langle-1|~~;~~H_R = \sum_{\vec{k}} \omega_k a^\dagger_{\vec{k} }a_{\vec{k} }\,. \label{HAHRfo}\ee The interaction Hamiltonian in the interaction picture and in the rotating wave approximation  is given by
\be H_{AR}(t)= \sum_{\vk}\Bigg\{  g_{k}\, a^{\dagger}_{\vk}\Big[ |+1\rangle\langle A|\,e^{i(k-\Omega_+)t} +   |-1\rangle\langle A|\,e^{i(k-\Omega_-)t}\Big] + \mathrm{h.c.} \Bigg\} \label{HARfo}\ee where

\be  \Omega_{\pm}=E_A-E_{\pm}~~;~~ g_{k} = -i\sqrt{\frac{k}{2V}}~  {D} \,,\label{Dipfo}   \ee  here $V$ is the volume and  $D$ is the dipole matrix element neglecting polarization  degrees of freedom.

 Consider that at time $t=0$ the initial state is
 \be |\Psi(0)\rangle =  |A\rangle\ |0_\gamma\rangle \label{inistatefo} \ee where $|0_\gamma\rangle$ is the radiation vacuum state, and write in the interaction picture the time evolved state  as
 \be |\Psi(t)\rangle_I = C_A(t)|A\rangle |0_\gamma\rangle + \sum_{\vk} |1_{\vk}\rangle \Big[ {C}_{k,+}(t)\, |+1\rangle + {C} _{k,-}(t)\, |-1\rangle\Big] \,. \label{wwstatefo}\ee The coefficients obey the following equations (in obvious notation)
 \bea \dot{C}_A(t) & = &  -i \langle A;0_\gamma|H_{AR}(t)|1_{\vk}; +1\rangle \,{C} _{k,+}(t)-i \langle A;0_\gamma|H_{AR}(t)|1_{\vk};-1\rangle \,{C} _{k,-}(t) \label{dotcAfo}\\ \dot{C} _{k,\pm}(t) & = &
 -i \langle 1_{\vk}; \pm 1 |H_{AR}(t)|  A;0_\gamma\rangle \,{C} _{A}(t) \,. \label{dotCpmfo}    \eea  We solve this system of equations with the initial conditions
 \be C_A(0)=1 ~;~ {C} _{k,\pm}(0) =0 \,.  \label{inicondfo} \ee In the Wigner-Weisskopf approximation\cite{book1,ww} the coefficients are   given by\footnote{We neglect the contribution from the Lamb shift to the energy level $E_A$.}

\bea  C_A(t)  & = &  e^{-\frac{\Gamma}{2} t}\,, \label{Caoft}\\
  {C}_{k,\pm}(t) & = &   i g_{k}  \, ~\frac{\Big[1-e^{i\big(k-\Omega_\pm +i\Gamma/2\big)t} \Big]}{\Big(k-\Omega_\pm +i\frac{\Gamma}{2} \Big)} \,.\label{Cpmfo}
  \eea

    The level width $\Gamma$ is given by
\be \Gamma = \Gamma_+ +  \Gamma_- \label{partialtotal}\ee where the partial widths $\Gamma_\pm $ correspond to the spontaneous decay channels $|A;0_\gamma\rangle  \rightarrow |1_{\vk} \rangle |+1\rangle; |A\rangle  \rightarrow |1_{\vk}\rangle |-1\rangle$ respectively, namely
\be  \Gamma_\pm = 2\pi\, \sum_{\vk}   |\langle A|H_{AR}(0)|1_{\vk}; \pm 1\rangle|^2 \,\delta(k-\Omega_\pm  ) =  2\pi\, \sum_{\vk}   |g_{k}|^2 \,\delta(k-\Omega_\pm  ) = \frac{D^2\Omega^3_\pm}{2\pi}\label{gamapmpfo}  \ee

 In most experimental circumstances, the energy splitting is much smaller than the optical frequency of the transitions, namely $|\Omega_+ - \Omega_-| \ll \Omega_\pm$ in which case it is convenient to write
\be \Omega_\pm = \Omega \pm \frac{\Delta \omega}{2} ~~;~~ \Delta \omega \ll \Omega \label{freqs}\ee and to leading order in $\Delta \omega/\Omega$ it follows that
\be \Gamma_+ \simeq \Gamma_- \simeq \Gamma/2 \,.\label{gamaslead}\ee


In the experiment reported in ref.\cite{dutt}, it has been verified that the approximation (\ref{gamaslead}) is fulfilled in the setting of that experiment. In what follows we will assume that the relation (\ref{gamaslead}) holds unless otherwise stated.


We write the spin qubit-photon entangled part of the wavefunction (in the interaction picture) (\ref{wwstatefo}) as
\be |\Psi_{sp}(t)\rangle =  \frac{1}{\sqrt{2}}\Big[ |\sigma_1(t)\rangle |+1\rangle + |\sigma_2(t)\rangle |-1\rangle \Big] \label{quphot} \ee where the single photon wavepackets are given by
\be |\sigma_{1}(t)\rangle = \sqrt{2} \sum_{\vk} {C}_{k,+}(t)|1_{\vk}\rangle ~~;~~|\sigma_{2}(t)\rangle = \sqrt{2} \sum_{\vk} {C}_{k,-}(t)|1_{\vk}\rangle\label{photpack}\ee

\subsection{Normalization of  photon wavepackets:} The normalization and orthogonality of the single photon wave-packets is determined by the overlaps
\be  \langle \sigma_{a}(t)|\sigma_{b}(t)\rangle = 2 \sum_{\vk}    {C}_{\vk,b}(t)\,  {C}^*_{\vk,a}(t) ~~;~~a,b = 1,2 \,. \label{normas}\ee Consider the functions
\be \mathcal{G}_{\alpha}(\omega,t) = \frac{\Big[1-e^{i(\omega-\Omega_\alpha+i\frac{\Gamma}{2}) t} \Big]}{\Big[\omega-\Omega_\alpha+ i\,\frac{\Gamma}{2} \Big]}\,  \label{funGs}\ee  in the narrow width limit $\Gamma \ll \Omega_{\alpha}$ these   are sharply localized near $\omega \simeq \Omega_\alpha$,   straightforward contour integration yields
\be \int_{-\infty}^{\infty} \mathcal{G}_{\alpha}(\omega,t)\mathcal{G}^*_{\beta}(\omega,t)\,d\omega = 2\pi \, \frac{\Big[1-e^{-i(\Omega_\alpha-\Omega_\beta)t}\,e^{-\Gamma t} \Big]}{\Gamma+i(\Omega_\alpha-\Omega_\beta)} \,. \label{inteGs}\ee

 Combining this result with (\ref{photpack},\ref{Cpmfo}) we find consistently with the Wigner-Weisskopf approximation
\be \langle \sigma_{1,2}(t)|\sigma_{1,2}(t)\rangle  = \frac{2\Gamma_{+,-}}{\Gamma}  \Big[1- e^{-\Gamma t} \Big]\,. \label{normasfin}\ee This result, along   with the relation between the total and partial decay widths given by (\ref{partialtotal}) yields the normalization of the $|\Psi_{sp}\rangle $ state,
\be  \langle \Psi_{sp}(t) | \Psi_{sp}(t)\rangle  = \Big[1-e^{-\Gamma t}  \Big] \label{unitary}\,, \ee which is a result of unitary time evolution manifest in the Weisskopf-Wigner formulation since the total state $|\Psi(t)\rangle_I$ given by (\ref{wwstatefo}) must obey $ \langle \Psi(t)|\Psi(t)\rangle  =1$. Because $|\sigma_{1,2}\rangle$ are single photon wavepackets, it is straightforward to confirm that the total number of photons is given by
\be N_\gamma(t) =  \langle \Psi_{sp}(t)\Big|\sum_{\vec{k}}a^\dagger_{\vec{k}}a_{\vec{k}}\Big|\Psi_{sp}(t)\rangle  =\Big[1-e^{-\Gamma t}  \Big]\,. \label{photonumbfo}\ee

Taking $\Gamma_+ \simeq \Gamma_- \simeq \Gamma/2$ under the assumption that $\Delta \omega \ll \Omega$, consistent with the experimental setup in \cite{dutt},  it follows that the single photon wavepackets are normalized for $\Gamma t \gg 1$ but they are \emph{not} orthogonal, we find
\be \langle \sigma_{2}(t)|\sigma_{1}(t)\rangle = \frac{\Big[1-e^{-i\Delta \omega t}\,e^{-\Gamma t} \Big]}{1+i\frac{\Delta \omega}{\Gamma}}\,,\label{overlap}\ee a result that is in agreement with an observation in ref.\cite{sham1} for $\Gamma t\gg 1$.

Let us  consider the reduced density matrix for the qubit by tracing over the radiation field, namely (in the interaction picture)
\be \rho^{I}_{fo}(t) = \mathrm{Tr}_{R} |\Psi_{sp}(t)\rangle \langle \Psi_{sp}(t) | \,,\label{redqubit}\ee going back to the Schroedinger picture we find
\be \rho_{fo}(t)= \frac{1}{2}\Big[1-e^{-\Gamma t}  \Big]\Bigg\{|+1\rangle \langle +1|+|-1\rangle \langle -1| + \Bigg(e^{i\Delta \omega t}\,\eta(t)\,
|+1\rangle \langle -1|   +\mathrm{h.c.}\Bigg) \Bigg\}  \label{redqubit2}\ee where
\be \eta(t) \equiv |\eta(t)|e^{i\varphi(t)} = \frac{\,\Big[1-e^{-i\Delta \omega \, t}~e^{-\Gamma t} \Big]}{\big(1-e^{-\Gamma t}  \big)\big(1+i\frac{\Delta \omega}{\Gamma}\big)}\,.\label{etadef}\ee

In the long time limit $\Gamma t \gg 1$ the coherence is suppressed by the factor $1/\sqrt{1+\Delta \omega^2/\Gamma^2}$ reflecting the   suppression of coherence by ``which-path'' information. If $\Gamma \gg \Delta \omega$ the spectral width of the radiation, determined by the lifetime of the excited state, suppresses which path information by blurring the energy resolution of the  decay channels of the  emitted photons   and coherence is maintained. In the opposite limit $\Delta \omega \gg \Gamma$ the energy difference between the lower lying states is resolved and which path information is available in the emission spectrum thereby suppressing coherence. This is manifest in the overlap of the photon wavepackets (\ref{overlap}) in terms of the product of the Lorentzian line shapes for the individual channels.

The main reason for studying the reduced density matrix in the case of frequency entanglement only is that, as it will be discussed in detail in section (\ref{sec:photo}) photodetection that filters horizontal (H) or vertically (V) polarized photons projects the density matrix onto a reduced density matrix precisely of the form (\ref{redqubit2}) that contains ``which-path'' information.

\subsection{Entanglement in frequency and polarization}

\begin{figure}[h!]
\includegraphics[keepaspectratio=true,width=2in,height=2in]{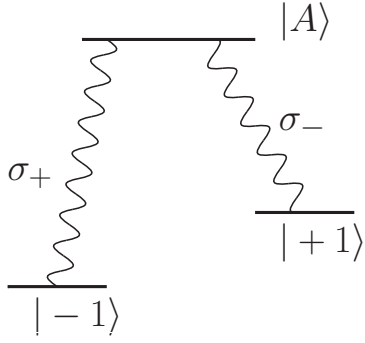}
\caption{Transitions  }
\label{fig:spent}
\end{figure}

In the experimental situations considered in refs.\cite{photatom1,photatom2,photatom3} for atom-photon entanglement and in ref.\cite{dutt} for electron spin-photon entanglement in NV-centers,   there are  angular momentum selection rules in spontaneous decay and the photons emitted are right handed (for $|A\rangle \rightarrow |-1\rangle$) or left handed (for $|A\rangle \rightarrow | + 1\rangle$) circularly polarized as depicted in fig. (\ref{fig:spent}). In this case the spin- qubit and the spontaneously emitted photons are entangled both in polarization and frequency.  Including the polarization of the emitted photons leads to  several important modifications of the results obtained in the previous case, therefore we restore the polarization, momentum and spatial dependence of the dipole matrix elements.  Although we focus the discussion on the experimental setup of ref.\cite{dutt} with NV-centers,   the results will be  more general.

In this case the total Hamiltonian for the three level $\Lambda$-system interacting with the electromagnetic field is given by (\ref{totalH}) with $H_A$ given in eqn. (\ref{HAHRfo}), but now
\be  H_R = \sum_{\vec{k},\lambda=\pm} \omega_k a^\dagger_{\vec{k},\lambda}a_{\vec{k},\lambda} \,,\label{HAR}\ee and the interaction Hamiltonian in the interaction picture and in the rotating wave approximation  is given by
\be H_{AR}(t)= \sum_{\vk}\Bigg\{  g_{\vk,+}(\vec{x}_0)\, a^{\dagger}_{\vk,-} |+1\rangle\langle A|\,e^{i(k-\Omega_+)t} +   g_{\vk,-} (\vec{x}_0)\, a^{\dagger}_{\vk,+} |-1\rangle\langle A|\,e^{i(k-\Omega_-)t} + \mathrm{h.c.} \Bigg\} \label{HAR2}\ee where $\Omega_{\pm}=E_A-E_{\pm}$, and

\be   g_{\vk,\pm}(\vec{x}_0) = -i\sqrt{\frac{k}{2V}}\, \vec{D}_{\pm}\cdot \vec{\epsilon}_{\vk,\mp}\,e^{i\vk\cdot \vec{x}_0} \label{Dpm}   \ee  here $V$ is the volume,  $\vec{D}_{\pm}$ are the dipole matrix elements $\langle \pm 1 |\vec{d}|A\rangle  $ respectively,  $\vec{\epsilon}_{\vk,\mp}$ are the left and right handed polarization vectors respectively and $\vec{x}_0$ is the position of the NV center.

 Consider that at time $t=0$ the initial state is
 \be |\Psi(0)\rangle =  |A\rangle\ |0_\gamma\rangle \label{inistate} \ee where $|0_\gamma\rangle$ is the radiation vacuum state, and following the notation of the previous section we write   the time evolved state  in the interaction picture  as
 \be |\Psi(t)\rangle_I = C_A(t)|A\rangle |0_\gamma\rangle + \sum_{\vk} \Bigg[ {C}_{\vk,+}(t)\,|1_{\vk,-}\rangle |+1\rangle + {C} _{\vk,-}(t)\,|1_{\vk,+}\rangle |-1\rangle\Bigg] \,. \label{wwstate}\ee The coefficients obey the following equations (in obvious notation)
 \bea \dot{C}_A(t) & = &  -i \langle A;0_\gamma|H_{AR}(t)|1_{\vk,+};-1\rangle \,{C} _{\vk,-}(t)-i \langle A;0_\gamma|H_{AR}(t)|1_{\vk,-};+1\rangle \,{C} _{\vk,+}(t) \label{dotcA}\\ \dot{C} _{\vk,\pm}(t)  & = &
 -i \langle 1_{\vk,\mp};\pm1 |H_{AR}(t)|  A;0_\gamma\rangle \,{C} _{A}(t) \,.\label{dotCpm}  \eea  Just as in the previous section we solve this system of equations with the initial conditions $
   C_A(0)=1 ~;~ {C} _{\vk,\pm}(0)  =0  $,  in the Wigner-Weisskopf approximation the coefficients are   given by\footnote{Again we neglect the contribution from the Lamb shift to the energy level $E_A$.}

\bea  C_A(t)  & = &  e^{-\frac{\Gamma}{2}\, t}\,, \label{Caoftfp}\\
  {C}_{\vk,\pm}(t) & = &   i g_{\vk,\pm}(\vec{x}_0) \, ~\frac{\Big[1-e^{i\big(k-\Omega_\pm +i\Gamma/2\big)t} \Big]}{\Big(k-\Omega_\pm +i\frac{\Gamma}{2} \Big)} \,.\label{Cpmfp}
  \eea

    The level width $\Gamma$ is given by
\be \Gamma = \Gamma_+ +  \Gamma_- \label{partialtotal2}\ee where the partial widths $\Gamma_+\,,\,  \Gamma_-$ correspond to the spontaneous decay channels $|A;0_\gamma\rangle  \rightarrow |1_{\vk,+} \rangle |-1\rangle; |A\rangle  \rightarrow |1_{\vk,-}\rangle |+1\rangle$ respectively, namely
\bea \Gamma_+ = 2\pi\, \sum_{\vk}   |\langle A|H_{AR}(0)|1_{\vk,-};+1\rangle|^2 \,\delta(k-\Omega_+  ) =  2\pi\, \sum_{\vk}   |g_{\vk,+}(\vec{x}_0)|^2 \,\delta(k-\Omega_+  ) \label{gamaplu} \\ \Gamma_- = 2\pi\, \sum_{\vk}   |\langle A|H_{AR}(0)|1_{\vk,+};+1\rangle|^2 \,\delta(k-\Omega_-  )  =  2\pi\, \sum_{\vk}   |g_{\vk,-}(\vec{x}_0)|^2 \,\delta(k-\Omega_- )\,. \label{gamamin}\eea Just as in the previous case of unpolarized photons, it follows that $\Gamma_\pm \propto |\vec{D}_\pm|^2\,\Omega^3_\pm$ but the proportionality constants now depend on the angular average of the polarization vectors.

Now the second term of the wave function (\ref{wwstate}) describes an entangled state of \emph{circularly polarized photons}  and the spin states of the NV- center, following the literature\cite{photatom1,photatom2,photatom3,dutt} we write this second term (in the interaction picture)  as
\be |\Psi_{sp}(t)\rangle  = \frac{1}{\sqrt{2}}\Big[  |\sigma_-(t)\rangle\, |+1\rangle +  |\sigma_+(t)\rangle\, |-1\rangle \Big] \label{entanstate}\ee where
\be |\sigma_\mp(t)\rangle   =   \sqrt{2}\,\sum_{\vk}   {C} _{\vk,\pm}(t)\,|1_{\vk,\mp}\rangle \label{sigminplus} \ee describe orthogonal circularly polarized single photon \emph{wave packets}.

Unlike the results in ref.\cite{sham1} we do not take the limit $\Gamma t \gg 1$, in the experimental setting of ref.\cite{dutt} the lifetime of the excited state is $1/\Gamma \approx 12 \,\mathrm{ns}$ but the measurements are performed during a time interval $\simeq 10-20 \,\mathrm{ns}$.

Borrowing the results from the previous section, we now find
\be    \langle \sigma_{\mp}(t)|\sigma_{\mp}(t)\rangle   =    \frac{2\,\Gamma _\pm}{\Gamma}\Big[1-e^{-\Gamma t}  \Big]~~;~~ \langle \sigma_{+}(t)|\sigma_{-}(t)\rangle =0 \,,\label{normaminplus}   \ee  where the orthogonality of  $|\sigma_{\mp}(t)\rangle$ is a consequence of the fact that they describe one photon wavepackets with  orthogonal polarizations.  This result, along   with the relation between the total and partial decay widths given by (\ref{partialtotal2}) again yields the normalization of the
$|\Psi_{sp}\rangle $ state,
\be  \langle \Psi_{sp}(t) | \Psi_{sp}(t)\rangle  = \Big[1-e^{-\Gamma t}  \Big] \label{unitary2}\,, \ee which is a result of unitary time   evolution and  similarly
\be N_\gamma(t) =  \langle \Psi_{sp}(t)\Big|\sum_{\vec{k},\lambda=\pm}a^\dagger_{\vec{k},\lambda}a_{\vec{k},\lambda}\Big|\Psi_{sp}(t)\rangle  =\Big[1-e^{-\Gamma t}  \Big]\,. \label{photonumbf}\ee

Just as in the previous section   the one-photon wave packets $|\sigma_{\pm}\rangle$ have unit normalization when $\Gamma t \gg 1$ \emph{and} $\Gamma_+=\Gamma_-=\Gamma/2$, which is   justified when the Zeeman splitting  $\Omega_+-\Omega_- \ll \Omega_+,\Omega_-$ and describes the experimental setup of ref.\cite{dutt}.
The reduced density matrix for the spin-qubit can be obtained by tracing over the radiation field just as in the previous section (\ref{redqubit},\ref{redqubit2}). However, in this case, the orthogonality of the circularly polarized wave packets leads to vanishing coherence and a diagonal density matrix  that describes a statistical mixture  given by
\be  \rho_{fp}(t)= \mathrm{Tr}|\Psi_{sp}(t)\rangle \langle \Psi_{sp}(t)| =  \frac{1}{2}\Big[1-e^{-\Gamma t}  \Big]\Big(|+1\rangle \langle +1|+|-1\rangle \langle -1| \Big)\,.\label{rhofp1}\ee

\subsection{Entanglement entropy:}
As we have seen above spontaneous generation of coherence leads to very different reduced density matrices depending on whether  photon-qubit entanglement is in frequency and polarization or   frequency only. This difference is highlighted by comparing the Von-Neumann entanglement entropy in both cases.

\vspace{2mm}

\textbf{Frequency entanglement only:} in this case the total reduced density matrix is
\be \rho(t) = e^{-\Gamma t} |A\rangle \langle A| + \rho_{fo}(t) \label{totfreqo}\ee where $\rho_{fo}(t)$ is given by (\ref{redqubit2}) which can be diagonalized with the following eigenvectors and eigenvalues
\bea \widetilde{|1\rangle}  & = & \frac{1}{\sqrt{2}}\Big( |+1\rangle + e^{-i\varphi(t)}\,e^{-i\Delta \omega t}|-1\rangle \Big)~~;~~ \lambda_1(t) =  \frac{1}{2}\Big[1-e^{-\Gamma t}  \Big]\,\Big[1+|\eta(t)|\Big]   \label{newvec1} \\
\widetilde{|2\rangle}  & = & \frac{1}{\sqrt{2}}\Big( |+1\rangle - e^{-i\varphi(t)}\,e^{-i\Delta \omega t}|-1\rangle \Big)~~;~~ \lambda_2 (t) =  \frac{1}{2}\Big[1-e^{-\Gamma t}  \Big]\,\Big[1-|\eta(t)|\Big]   \label{newvec2}\,\eea where $\eta(t)=|\eta(t)|e^{i\varphi(t)}$ is given by (\ref{etadef}), leading to
\be \rho(t) = e^{-\Gamma t} |A\rangle \langle A|+ \lambda_1(t)~ \widetilde{|1\rangle} \widetilde{\langle 1|}+\lambda_2(t)~ \widetilde{|2\rangle} \widetilde{\langle 2|} \,.\label{rhodiag}\ee The entanglement entropy follows directly,
\be S_{fo}(t) = \Gamma t e^{-\Gamma t}- \lambda_1(t) \ln \lambda_1(t)-\lambda_2(t) \ln \lambda_2(t) \,. \label{entropyfo}\ee For $\Gamma t \gg 1$
\be S_{fo}(\infty) = -\frac{1}{2}\bigg[ {1+|\eta_\infty|} \bigg]\,\ln\bigg[\frac{1+|\eta_\infty|}{2}\bigg]-
\frac{1}{2}\bigg[1-|\eta_\infty|\bigg]\,\ln\bigg[\frac{1-|\eta_\infty|}{2}\bigg] \label{sfoinfty}\ee with
\be |\eta_\infty| = \frac{1}{\sqrt{1+\frac{\Delta \omega^2}{\Gamma^2}}} \label{alfainfty}\ee As $\Delta \omega/\Gamma \rightarrow 0$ the entanglement entropy vanishes as the asymptotic state is the  pure state $\widetilde{|1\rangle}= \frac{1}{\sqrt{2}}\Big( |+\rangle + |-\rangle \Big)$ in the opposite limit $\Delta \omega /\Gamma \gg 1$ where which-path information suppresses coherence it follows that $S_{fo}(\infty) = \ln(2)$ describing an equal probability statistical mixture.

\vspace{2mm}

\textbf{Entanglement in frequency and polarization:} in this case   the total reduced density matrix is simply
\be \rho(t) = e^{-\Gamma t} |A\rangle \langle A|+ \frac{1}{2}\Big[1-e^{-\Gamma t}  \Big] \Big(|+1\rangle \langle +1 + |-1\rangle \langle -1|\Big)  \, \label{rhodiagpol}\ee as a consequence of the orthogonality of the right and left circular polarized photon wavepackets. In this case the   entanglement entropy is
\be S_{fp}(t)= \Gamma t e^{-\Gamma t}-[1-e^{-\Gamma t}]\ln\Big[\frac{1-e^{-\Gamma t}}{2} \Big] \label{Sfpoft}\ee with the asymptotic value
\be S_{fp}(\infty) = \ln(2) \,.\label{asySpol}\ee

\begin{figure}[h!]
\includegraphics[keepaspectratio=true,width=3.5 in,height=3.5 in]{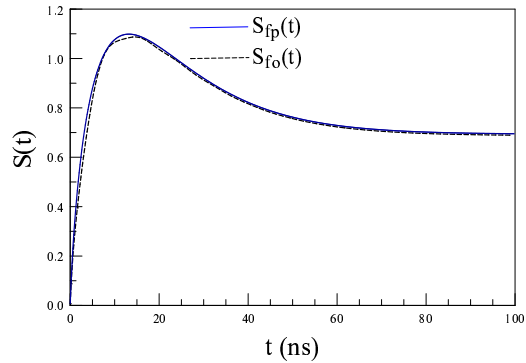}
\caption{Entanglement entropy for the case of entanglement in frequency only $S_{fo}(t)$ and frequency and polarization $S_{fp}(t)$ for $\Delta \omega = 2\pi\times 122\,\mathrm{MHz}~;~\Gamma = 1/12\,\mathrm{ns}, \Gamma_+ = \Gamma_- = \Gamma/2$.   }
\label{fig:entropy}
\end{figure}

The entanglement entropies in both cases are displayed in fig.(\ref{fig:entropy}) for the parameters of the experiment in ref.\cite{dutt}, $\Delta \omega = 2\pi\times 122\,\mathrm{MHz}~;~\Gamma = 1/12\,\mathrm{ns}$.

Analytically it can be seen that
\be S_{fp}(t) \geq S_{fo}(t)\,, \label{great}\ee a relation that is confirmed numerically
 and confirms the qualitative expectation that the entanglement entropy should be larger in the case of entanglement both in frequency and polarization.

The results above were obtained under the assumption that $\Gamma_+=\Gamma_-= \Gamma/2$.
If the partial widths to the two non-degenerate levels are different    the generalized form of the entanglement entropy in this case of entanglement in frequency and polarization is given by
\be S_{fp}(t)= \Gamma t e^{-\Gamma t}-\frac{\Gamma_+}{\Gamma}\big(1-e^{-\Gamma t}\big)\ln\Big[\frac{\Gamma_+}{\Gamma}\big(1-e^{-\Gamma t}\big) \Big]-\frac{\Gamma_-}{\Gamma}\big(1-e^{-\Gamma t}\big)\ln\Big[\frac{\Gamma_-}{\Gamma}\big(1-e^{-\Gamma t}\big)\Big] \label{Sfpoftgen}\ee where $\Gamma=\Gamma_+ + \Gamma_-$.

\section{Photodetection}\label{sec:photo} We consider a model for a broadband photodetector  described by an atom localized at position $\vec{x}_d$  interacting with the radiation field in the dipole approximation \emph{a l\'a} Glauber\cite{glauber,book1,book2}. The Hamiltonian is given by $H_D + H_{DR}$ where the detector Hamiltonian $H_D$ describes a zero energy ground state and a collection of excited states which eventually will be taken as a continuum
\be H_D = \nu_0 \, |g^d\rangle \langle g^d| + \sum_{j}\nu_j |e^d_j\rangle \langle e^d_j| ~~;~~\nu_0=0 \,, \label{Hdet}\ee and $H_{DR}$ is the interaction Hamiltonian that describes a dipolar coupling to the radiation field with a filter that selects $H/V$ linear polarization states of the radiation field. In the rotating wave approximation and in the interaction picture it is given by
\be H_{DR}(t) = \sum_{j}\Big[ \vec{d}_j \cdot \vec{E}^{(+)}_{P}(\vec{x}_d;t) |e^d_j\rangle\langle g^d|\,e^{i\nu_j t} + \mathrm{h.c.}\Big]~~;~~P=H;V \,\label{HDR}\ee $\vec{d}_j$ are the dipole matrix elements and
\be \vec{E}^{(+)}_{P}(\vec{x}_d;t) = \sum_{\vec{k}} i\sqrt{\frac{k}{2V}}~\vec{\epsilon}_P~a_{\vec{k},P}\,e^{i\vec{k}\cdot{\vec{x}_d}}e^{-ikt} \,.\label{Efield}\ee The combined process of spontaneous emission from the NV-center $|A\rangle$ considered to be localized at $\vec{x}_0 = \vec{0}$ and photodetection by a broadband photodetector localized at $\vec{x}_d$ is now described by the \emph{total} Hamiltonian
\be H_{tot} = H_A +  H_R + H_{AR}+H_D +H_{DR}\,, \label{Htot}\ee where the first three terms are given by (\ref{totalH}-\ref{HARfo}).

Insight into the combined processes and the intermediate states that contribute is gleaned in second order in the perturbative expansion with the full interaction Hamiltonian in the interaction picture (and in the rotating wave approximation)
\be H_I(t) = H_{AR}(t)+ H_{DR}(t) \label{HIful}\ee where $H_{AR}(t);H_{DR}(t)$ are given by (\ref{HAR2}) with $\vec{x}_0=\vec{0}$ and (\ref{HDR}) respectively. Consider that the initial state is (in obvious notation)
\be |\Psi(0)\rangle = |A;0_\gamma;g^d\rangle \,,\label{psitotini}\ee in the interaction picture the resulting time dependent state in second order becomes
\be |\Psi(t)\rangle = \Bigg[1-i\int^t_0 H_I(t_1)dt_1 +(-i)^2\int^t_0 \int^{t_{1}}_0 H_I(t_1)  H_I(t_2)dt_1 dt_2+\cdots \Bigg]|\Psi(0)\rangle \,.\label{psi2nd}\ee To first order only $H_{AR}$ contributes and describes the perturbative spontaneous decay of the excited state $|A\rangle$ of the NV-center into the Zeeman split states $|1_{\vec{k},+};- 1\rangle$ and   $|1_{\vec{k},-};+ 1\rangle$. Inserting a complete set of eigenstates of $H_0=H_A+H_D+H_R$ it is straightforward to see that in the second order contribution the first term $H_I(t_2)$ describes the spontaneous emission of the circularly polarized photons  while the second term $H_I(t_1)$ describes the absorption of these photons and the photoexcitation of the detector (along with a second order contribution from $H_{AR}$ that yields the original state back).
The photodetection probability at time $t$ is given by\cite{glauber,book1,book2}
\be P_D(t) = {\mathrm{Tr}}_{d} \sum_{j}|e^d_j\rangle\langle e^d_j| \rho(t) \,,\label{detprob}\ee where the density matrix
\be \rho(t) = |\Psi(t)\rangle \langle \Psi(t)| \,,\label{rho}\ee and the trace in (\ref{detprob}) is over the detector excited states.

Our goal is to describe these processes \emph{non-perturbatively} with a Wigner-Weisskopf description that incorporates both processes at once. Guided by this perturbative analysis, we propose the following form of the time dependent state in the interaction picture
\be |\Psi(t)\rangle = |\Psi_A(t)\rangle \,|g^d \rangle + |\Psi_{DS}(t)\rangle \,|0_\gamma \rangle \label{psioft}\ee where
\be |\Psi_A(t)\rangle = C_A(t)|A\rangle |0_\gamma\rangle + \sum_{\vk} \Bigg[ {C}_{\vk,+}(t)\,|1_{\vk,-}\rangle |+1\rangle + {C} _{\vk,-}(t)\,|1_{\vk,+}\rangle |-1\rangle\Bigg] \label{psiAoft}\ee and
\be |\Psi_{DS}(t)\rangle = \sum_{j}\Big[D_{j,-}(t)|-1\rangle + D_{j,+}(t)|+1\rangle \Big]|e^d_j\rangle \,, \label{psiDAoft}\ee with the initial conditions
\be C_A(0) =1~;~C_{\vec{k},\pm}(0) =0 ~;~D_{j,\pm}(0) = 0 \,.\label{inis}\ee
We highlight that $|\Psi_{DS}(t)\rangle$ describes an entangled state between the spins and the detector.

The explicit solution for the coefficients with the initial conditions (\ref{inis}) is provided in the appendix.

 The coefficients $D_{j,\pm}(t)$  determine the photodetection probability and display the causal nature of the propagation\cite{causality}: the detection time $t_D$ has to be larger than $t_d = x_d/c$, namely the time it takes the front of the photon pulse to travel from the NV-center to the position of the photodetector.  In the experimental setup of ref.\cite{dutt} the photon travels along a $\sim 2\,\mathrm{m}$ long fiber to the photodetector, therefore $t_d \simeq 7\,\mathrm{ns}$.

 The   photodetection probability is obtained as in (\ref{detprob}), and obviously only the state $|\Psi_{DA}(t)\rangle$ contributes. The result is a \emph{projected} reduced density matrix for the spin-qubit subpace $|\pm 1\rangle $
 namely
 \bea \rho^{(I)}_D(t) = {\mathrm{Tr}}_{d} \sum_{j}|e^d_j\rangle\langle e^d_j| |\Psi_{DS}(t)\rangle\langle \Psi_{DS}(t)| = && \sum_{j} \Bigg[|D_{j,+}(t)|^2 ~|+1\rangle\langle +1| +  |D_{j,-}(t)|^2 ~|-1\rangle\langle -1| \nonumber \\ +&& \Big( D_{j,+}(t)D^*_{j,-}(t)~|+1\rangle\langle -1|+\mathrm{h.c.}\Big) \Bigg]\,. \label{reddensmtx}\eea where the coefficients $D_{j,\pm}(t)$ are given in the appendix by (\ref{djotasfin}). We now introduce the density of states of the photodetector $\mathcal{D}(\omega)$: for any arbitrary function of the detector frequencies $\mathcal{F}(\nu_j)$
 \be \sum_j |\kappa^2_j|\,\mathcal{F}(\nu_j) = \int_{-\infty}^\infty
  \mathcal{D}(\omega) \mathcal{F}(\omega) d\omega ~~;~~\mathcal{D}(\omega) = \sum_{j} |\kappa^2_j|\delta(\omega-\nu_j) \,.\label{densofstate}\ee With the result for $D_{j,\pm}(t)$ given in the appendix (\ref{djotasfin}), we introduce  \be \mathcal{F}_{\pm}(\omega;t) = \sqrt{\frac{\Gamma_\mp}{2\pi}}~ \frac{\Big[1-e^{i(\omega-\Omega_\pm+i\frac{\Gamma}{2}) (t-t_d)} \Big]}{\Big[\omega-\Omega_\pm + i\,\frac{\Gamma}{2} \Big]}\, \label{Fpm}\ee in terms of which the projected reduced density matrix  at the \emph{photodetection time} $t_D$ in the interaction picture becomes
  \bea \rho^{(I)}_D(t_D) =   \int^{\infty}_{-\infty}&&\mathcal{D}(\omega) \Big\{|\mathcal{F}_{+}(\omega;t_D)|^2~|+1\rangle\langle +1| + |\mathcal{F}_{-}(\omega;t_D)|^2~|-1\rangle\langle -1|\nonumber \\ + &&    \delta^P_- ~\mathcal{F}_{+}(\omega;t_D) \mathcal{F}^*_{-}(\omega;t_D) |+1\rangle\langle -1|+\mathrm{h.c.}\Big\} d\omega ~\Theta(t_D-t_d)\,.\label{densa}\eea   In the narrow width limit $\Gamma \ll \Omega_{\pm}$ the functions $\mathcal{F}_{\pm}(\omega;t)$ feature sharp peaks at $\omega = \Omega_\pm= \Omega \pm \Delta \omega/2$,  again we assume that $\Delta \omega \ll \Omega$ and consequently that $\Gamma_+ \simeq \Gamma_-\simeq \Gamma/2$. We also assume a broadband detector whose  spectral density   is insensitive to the spectral width of the emitted photon $\Gamma$ and the energy difference between the $|\pm 1\rangle$ states  $\Delta \omega$, namely  $\mathcal{D}(\Omega_\pm) \simeq \mathcal{D}(\Omega)$.  In particular the correlation function for the broadband photodetector\cite{book1,book2} is given by
  \be G_D(t-t')= \sum_{j} |\vec{d}_j|^2 e^{i\nu_j(t-t')} \propto \int_{-\infty}^{\infty} \mathcal{D}(\omega)\,e^{i\omega(t-t')}\, d\omega \sim  2\pi\,\mathcal{D}(\Omega) \, \delta(t-t') \,.\label{broadband}\ee

  We can now extract $\mathcal{D}(\omega) \simeq \mathcal{D}(\Omega)$ outside  the integrals, and using the result (\ref{inteGs}) we find
  \be \int^\infty_{-\infty} \mathcal{F}_{a}(\omega;t_D) \mathcal{F}^*_{b}(\omega;t_D) d\omega = \frac{ \sqrt{\Gamma_a\Gamma_b}}{ \Gamma}~\frac{\Big[1-e^{-i\Delta_{ab}(t_D-t_d)}\,e^{-\Gamma(t_D-t_d)} \Big]}{1+i\frac{\Delta_{ab}}{\Gamma}}~~;~~\Delta_{ab} = \Omega_a -\Omega_b ~~; a,b = +,- \label{Fintegs}\ee Going back to the Schroedinger picture at time $t_D$ and taking $\Gamma_+ = \Gamma_- = \Gamma/2$  the final result for the projected reduced density matrix is given by
  \bea \rho_D(t_D) & = &   \frac{\mathcal{D}(\Omega)}{2}\Big[1-e^{-\Gamma \tau}\Big]~\Theta(\tau)~\Bigg\{|+1\rangle\langle +1|+|-1\rangle\langle -1|\nonumber \\
  &+&    \delta^P_{-}~ |+1\rangle\langle -1|\,e^{i\Delta \omega \, t_D}\,\eta(\tau)+\mathrm{h.c.} \Bigg\}~~;~~\tau=t_D-t_d\label{findensa}\eea where $\eta(\tau)$ is given by (\ref{etadef}) with $\tau = t_D-t_d$.

  Comparing the prefactor of this expression with the total photon number (\ref{photonumbf}) it is clear that the prefactor is just describing the number of photons detected at the retarded time $t_D-t_d$ and allows the identification of $\mathcal{D}(\Omega)$ with the detection efficiency. In the experimental setup in\cite{dutt} this efficiency is $\ll 1$ thus justifying the neglect of the photon emission from the decay of the excited states of the detector. The coherence term has a simple interpretation:   photodetection by  filtering the linear polarizations $H$ or $V$    projects the spin-qubit-photon entangled state at a time $t_D$ into a state similar to that studied in section (\ref{subsec:freqonly}) effectively disentangling the polarization from the spin degree of freedom    leaving frequency entanglement only. For $\Gamma \tau \gg 1$ the coherence is suppressed by the same factor as in the previous case (\ref{redqubit2})  reflecting which path information.

  This result is fully compatible with Glauber's theory of photodetection with an ``ideal'' broadband photodetector\cite{glauber,book1,book2}, where the detection probability is given by
  \be P_D(t_D) = \kappa \int^{t_D}_0 \langle \sigma_\pm(t)|E^{(-)}(\vec{x_d},t)E^{(+)}(\vec{x_d},t)| \sigma_\pm(t)\rangle \,dt =\kappa' \, \frac{\Gamma_{\mp}}{\Gamma} \Big[1-e^{-\Gamma \tau}\Big]~\Theta(\tau)\label{glauber}\ee here $\kappa,\kappa'$ are constants\cite{book1} and we used eqn. (\ref{mtxele}). Similarly the interference terms are given by
  \bea P_I(t_D) = \kappa \int^{t_D}_0 \langle \sigma_+(t)|E^{(-)}(\vec{x_d},t)E^{(+)}(\vec{x_d},t)| \sigma_-(t)\rangle \, dt= &&\kappa'\, \frac{\sqrt{\Gamma_+\Gamma_-}}{\Gamma} \frac{\Big[1-e^{-i\Delta \omega\tau}\,e^{-\Gamma \tau}\Big]}{\Big[1+i\frac{\Delta \omega}{\Gamma} \Big] }\nonumber \\&&\times \delta^P_-~\Theta(\tau)\label{glauberI}\eea These are precisely the terms in the reduced density matrix (\ref{findensa}).

\begin{figure}[h!]
\includegraphics[keepaspectratio=true,width=3 in,height=3 in]{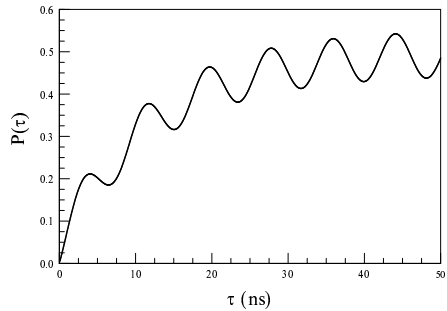}
\caption{The probability (\ref{finprobi}) for $t_d = 7 \,\mathrm{ns}~,~\Delta \omega = 2\pi\times 122\,\mathrm{MHz}~;~ 1/\Gamma = 12 \, \mathrm{ns}$.  }
\label{fig:proba}
\end{figure}

    After projection of the photon state into $H/V$ polarization,  spin-qubit-photon entanglement is displayed by projecting on any state of the form
  \be |M\rangle = \frac{1}{\sqrt{2}}\big[ |+1\rangle + e^{i \phi}|-1\rangle\big]\,. \label{magstate}\ee   This is implemented with the reduced density matrix (\ref{findensa}) by obtaining the conditional probability
  \be P_{M|H,V}(t_D) = Tr \rho_D(t_D) |M\rangle\langle M| \,.\label{pmhv}\ee The non-vanishing coherence in (\ref{findensa}) in the basis $|\pm 1\rangle$ leads to oscillatory behavior of $P_{M|H,V}(t_D)$ as a function of $t_D$.  For the state (\ref{magstate}) with $\phi =0$ and an $H$ projection  we find for $\tau = t_D-t_d >0$
  \be \frac{ {P}_{M|H}(\tau)}{\mathcal{D}(\Omega)} = \frac{1}{2} \big[1-e^{-\Gamma \tau}\big]\,\Bigg[1 + \mathrm{Re}~\Big( e^{i\Delta \omega \, t_D}\,\eta(\tau) \Big) \Bigg] ~~;~~\tau = t_D-t_d \label{finprobi}\ee
Fig. (\ref{fig:proba}) displays the probability (\ref{finprobi}) as a function of $\tau = t_D-t_d$ for the experimental values reported in ref.\cite{dutt}: $\Delta \omega = 2\pi\times 122\,\mathrm{MHz}~;~ 1/\Gamma = 12 \, \mathrm{ns} ~~;~~ t_d = 7 \,\mathrm{ns}$.

This figure reveals the effect of which-path suppression of the coherence: the asymptotic behavior of the probability is
 \be \frac{ {P}_{M|H}(\tau \gg 1/\Gamma)}{\mathcal{D}(\Omega)} \simeq \frac{1}{2}\Big[1+ \frac{\Gamma}{\Delta \omega}\,\sin\big[\Delta \omega (\tau+t_d)\big]\Big]\,. \label{asyprobawp}\ee Measurement in the $H/V$ basis results in a post-measurement density matrix that features coherence in the qubit basis $|\pm 1 \rangle$ suppressed by which-path information. This coherence was not manifest in the pre-measurement density matrix because of the orthogonality of the circularly polarized photon wave packets.

 The   reduced density matrix (\ref{findensa}) is similar to (\ref{redqubit2}), normalizing   so that $\widetilde{\rho}_D(\tau) = \rho_D(\tau)/\mathrm{Tr}\rho_D(\tau)$  it can be diagonalized in a new basis that differs from (\ref{newvec1},\ref{newvec2}) by the phases multiplying $|-1\rangle$ and with eigenvalues
  \be \epsilon_{\pm}(\tau) = \frac{1}{2}\Big[1\pm |\eta(\tau)|\Big] \label{epsilons}\ee
  respectively, leading to the  \emph{post-photodetection} Von-Neumann entropy of entanglement
 \be \widetilde{S}_D(\tau)  = -\mathrm{Tr}\widetilde{\rho}_D(\tau)\ln \widetilde{\rho}_D(\tau)=-\epsilon_+(\tau)   \ln \epsilon_+(\tau) -\epsilon_-(\tau)   \ln \epsilon_-(\tau)\,.  \label{Sdetec}\ee This post-measurement entanglement entropy is given by $S_{fo}(\infty)$ in eqn. (\ref{sfoinfty}) asymptotically for $\Gamma \tau \gg 1$  .

  \subsection{Implementing a ``Quantum eraser'':}

 The factor  $1/(1+i\Delta \omega/\Gamma)$ in the results (\ref{findensa},\ref{glauberI},\ref{etadef}) reflects which-path information because it suppresses coherence when $\Delta \omega \gg \Gamma$. It is noteworthy that  this suppression   remains in the final expressions even in an ``ideal'' broadband photodetector \emph{a l\'a} Glauber which is insensitive to the photon frequency and with a photodetection correlation function   $\propto \delta(t-t')$ as discussed above.

  In the experiment in ref.\cite{dutt} $\Delta \omega = 2\pi \times 122\,\mathrm{MHz}~~;~~ \Gamma \simeq 1/12\,\mathrm{ns} $ so that $\Delta \omega/\Gamma \simeq 9.2$ and there is a strong suppression of coherence because of which-path information  $1/\sqrt{1+\Delta \omega^2/\Gamma^2}\simeq 0.11$. In this experiment
   photodetection is carried out with a photodetector   with time resolution $\delta t \simeq 300\,\mathrm{ps} \ll 1/\Delta \omega $ to implement a ``quantum eraser''\cite{eraser1,eraser2} to ``erase'' which-path information by introducing an energy uncertainty $\sim 1/\delta t \gg \Delta \omega$.

    A simple model for such photodetector can be implemented by modifying the   interaction Hamiltonian between the detector and the radiation field $H_{DR}$ (\ref{Hdet})   introducing a ``\emph{shutter function}'' $\mathds{S}(t)$  with explicit time dependence, namely
  \be H_{DR}(t) =    \sum_{j}\Big[ \vec{d}_j \cdot \vec{E}^{(+)}_{P}(\vec{x}_d;t) |e^d_j\rangle\langle g^d|\,e^{i\nu_j t} + \mathrm{h.c.}\Big]\,\mathds{S}(t)~~;~~P=H;V \,, \label{shutterH} \ee where the \emph{only} restrictions on the shutter function $\mathds{S}(t)$ are
  \be \mathds{S}(t) = \Bigg\{ \begin{array}{l}
                       \sim 1 ~~   t_D -\delta t \leq t \leq t_D  \\
                       0~~\mathrm{otherwise}
                     \end{array} \label{delt} \ee with the shutter interval $\delta t$ such that
\be    \Gamma \delta t \ll \Delta \omega \delta t \ll 1\,. \label{indt}\ee     This function effectively describes a shutter with a time resolution $\delta t$ and amounts to ``slicing'' or time-binning the photon wavefunction upon detection.

   A similar procedure of ``chopping'' the wave function in short time intervals has also been advocated as a quantum eraser in ref.\cite{sham1}. In ref.\cite{mismatch} a phenomenological damping term is added to the right hand side of the equivalent of equations (\ref{dotDp}) in this reference, with the argument that such damping term describes the coupling of the (single) excited state of the detector atom to some reservoir. A ``quantum eraser'' is implemented in this approach by taking the damping constant   $\gamma \gg \Delta \omega$. While this phenomenological approach seems sensible, we consider instead    the model of the photodetector with the shutter function $\mathds{S}(t)$ introduced above implemented within an ideal broadband photodetector as follows.

   The solution for the coefficients $D_{j,\pm}(t_D)$ are now given by
   \be D_{j,\pm}(t_D)  = -i \frac{\vec{d}_j}{\sqrt{2}}\cdot \int_0^{t_D}    \langle 0_\gamma | \vec{E}^{(+)}_P(\vec{x}_d,t)|\sigma_\mp(t)\rangle \,  \mathds{S}(t)\,\,e^{i\nu_j t}   dt\,,\label{newds} \ee and the reduced density matrix elements in (\ref{reddensmtx}) become
   \bea \sum_{j} D_{j,a}(t_D) D^*_{j,b}(t_D) & = &  \frac{1}{2} \int_0^{t_D} dt  \int_0^{t_D}   dt' ~ \mathds{S}(t)\,\mathds{S}(t') \langle \sigma_b(t)|E^{(-)}(\vec{x_d},t)E^{(+)}(\vec{x_d},t')| \sigma_a(t')\rangle \nonumber \\ &\times & \int_{-\infty}^{\infty} \mathcal{D}(\omega) e^{i\omega(t-t')} \,d\omega ~~;~~ a,b = +,-\,, \label{newDcorr}\eea where we have used that $|\sigma_\pm(t)\rangle$ are one-photon wavepackets and only the vacuum contributes to the intermediate state in the correlation function of the electric field. The last term in (\ref{newDcorr}) is the photodetector correlation function \cite{book1,book2} which for a broadband photodetector is given by eqn. (\ref{broadband}), leading to
   \bea \sum_{j} D_{j,a}(t_D) D^*_{j,b}(t_D)  & = &  2\pi\,\frac{\mathcal{D}(\Omega)}{2} \int_0^{t_D} dt   \mathds{S}^2(t)\, \langle \sigma_b(t)|E^{(-)}_P(\vec{x_d},t)E^{(+)}_P(\vec{x_d},t)| \sigma_a(t)\rangle \nonumber \\ & \simeq &  2\pi\, \frac{\mathcal{D}(\Omega)}{2} \langle \sigma_b(t_D)|E^{(-)}_P(\vec{x_d},t_D)E^{(+)}_P(\vec{x_d},t_D)| \sigma_a(t_D)\rangle \, \delta t \,.\label{shutterfin}\eea  where we have used the condition (\ref{indt}) so that the integrand is constant in the interval $t_D-\delta t \leq t \leq t_D$ and vanishes outside it. Using the result (\ref{mtxele}) we obtain the reduced density matrix in the Schroedinger picture
    \be  \rho_D(t_D)   =     \frac{\mathcal{D}(\Omega)}{2}\,\Big(\Gamma \delta t\Big)\,\,e^{-\Gamma\tau} ~\Theta(\tau)~\Bigg\{|+1\rangle\langle +1|+|-1\rangle\langle -1|+  \delta^P_{-}~ \Big(|+1\rangle\langle -1|\, e^{i\Delta\omega \,t_d}  +\mathrm{h.c.} \Big) \Bigg\}\,.\label{findensashutter}\ee

    Remarkably, this density matrix describes a pure state, namely
    \be \rho_D(t_D) =  \mathcal{N}(\tau)~ \Big( e^{i\Omega_+t_d}|+1\rangle + \delta^P_{-}\,e^{i\Omega_-t_d}|-1\rangle\Big)\Big( e^{-i\Omega_+t_d}\langle+1| + \delta^P_{-}\, e^{-i\Omega_-t_d}\langle-1|\Big)\label{purstate}\ee with the normalization
    \be \mathcal{N}(\tau) = \frac{\mathcal{D}(\Omega)}{2}\,\Big(\Gamma \delta t\Big)\,\,e^{-\Gamma\tau} ~\Theta(\tau) ~~;~~\tau = t_D-t_d \, .\label{norma}\ee

It is noteworthy that the quantum eraser has \emph{purified} the post-measurement reduced density matrix.
This analysis confirms the experimental results in refs.\cite{dutt,qdotyama} and bolsters the arguments presented in ref.\cite{qdotyama}.

\begin{figure}[h!]
\includegraphics[keepaspectratio=true,width=2 in,height=2 in]{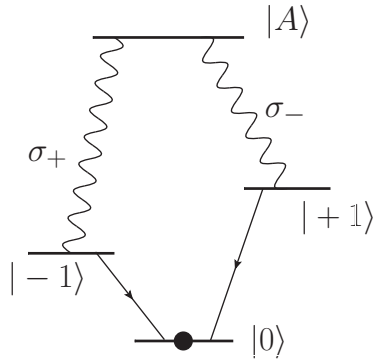}
\caption{Coherent transfer of the state $|M(t)\rangle$ to the ground state $|0\rangle$ see ref.\cite{dutt}. }
\label{fig:projection}
\end{figure}

In the experiment in ref.\cite{dutt} after detection the spin-qubit evolves freely in time from $t_D$ until
a time $t$ so that
\be \rho_D(t ) = \mathcal{N}(\tau)~ \Big( e^{i\Omega_+(t-\tau)}|+1\rangle + \delta^P_{-}\,e^{i\Omega_-(t-\tau)}|-1\rangle\Big)\Big( e^{-i\Omega_+(t-\tau)}\langle+1| + \delta^P_{-}\, e^{-i\Omega_-(t-\tau)}\langle-1|\Big)\label{purstatet}\ee
at which time two microwave pulses resonant with the levels $|\pm\rangle$ are turned on and transfer coherently the  state
\be |M(t)\rangle = \frac{1}{\sqrt{2}}\Big(e^{i\Omega_+ t}|+1\rangle + e^{i\Omega_- t}e^{i\phi}|-1\rangle \Big) \label{microM}\ee with a fixed phase $\phi$   to the ground state $|0\rangle$, as depicted in fig. (\ref{fig:projection}).
Now we find the total (joint) probability
\be P_{M|H,V}(\tau) = Tr \Big[\rho_D(t) |M(t)\rangle\langle M(t)|\Big] = \frac{\mathcal{N}(\tau)}{2}~\Big[1\pm \cos \alpha(\tau) \Big]~~;~~\alpha(\tau) = \Delta \omega ~\tau + \phi \,.\label{pmhvshut}\ee

This result agrees with the joint probability quoted and experimentally confirmed  in ref.\cite{dutt} up to the overall normalization factor and the retardation in the detection time $\tau = t_D-t_d$.

\section{Summary   and conclusions:}

In this article we have studied the dynamics of frequency and polarization entanglement between photons and a spin-qubit from spontaneous decay in a typical $\Lambda$   system with non-degenerate lower levels. We addressed  in detail how which path information affects coherence, obtained the entanglement entropy for the reduced spin-qubit with frequency and polarization entanglement and provided a unified description of the process of spontaneous emission and broadband photodetection that is fully causal and allows to include a quantum eraser in a consistent manner.

The main results are the following: beginning with the case in which photon spin-qubit entanglement does not involve polarization but only frequency, the reduced qubit density matrix obtained from tracing out the radiation bath features oscillatory coherence terms (in the qubit basis) that are suppressed by which path information by a factor $1/\sqrt{1+\Delta \omega^2/\Gamma^2}$  where $\Delta \omega$ is the Zeeman splitting between the lower spin states and $\Gamma$ is the linewidth of the excited state. In the case in which the spin degree of freedom is entangled with circularly polarized photons, the reduced density matrix is a statistical mixture as a consequence of the orthogonality of the polarization of the photon states. We obtain the entanglement Von-Neumann entropy in both cases and analyze their long time asymptotic behavior. In the case in which the spontaneous decay rate is the same to the two lower levels, we find that $S_{fp}(t) \geq S_{fo}(t)$ where $S_{fp}(t)$ ($S_{fo}(t)$) is the entanglement entropy for frequency and polarization (frequency only). Focusing on broadband photodetection in the case of frequency and polarization entanglement, we find that with an ideal photodetector that filters photons with horizontal (H) or vertical (V) directions the post-measurement density matrix describes a mixed state with non-vanishing coherences in the qubit basis. Despite the broadband nature of the photodetector described by correlation function $\propto \delta(t-t')$,  the coherences display oscillatory behavior suppressed by which path information just as the pre-measurement density matrix in the case of frequency entanglement.

A   ``quantum eraser'' is implemented within the Glauber model of broadband photodetection by including a ``shutter function''  that effectively   time-bins photodetection with a time resolution $\delta t$ so that $\Gamma \delta t \ll \Delta \omega \delta t \ll 1$ thereby introducing enough energy uncertainty   to average out frequency information. We find that photodetection with this ``quantum eraser'' \emph{purifies} the   post-measurement reduced density matrix to a  \emph{pure state}. The resulting joint probability for $H/V$ photodetection   with projection onto a a superposition of qubit states $|M(t)\rangle = \frac{1}{\sqrt{2}}\Big(e^{i\Omega_+ t}|+1\rangle + e^{i\Omega_- t}e^{i\phi}|-1\rangle \Big) $  is given by (\ref{pmhvshut}) and  agrees with the experimental results found in ref.\cite{dutt}.

Several aspects of the results obtained in this article suggest possible experimental avenues: 1) the dependence on the delay time $t_d = x_d/c$ with $x_d$ the position of the photodetector, suggests the possibility of using several photodetectors in coincidence, for example to study interference effects, Hanbury-Brown-Twiss correlations or as a complementary variable to explore coherence as a function of this delay distance, 2) rather than implementing a ``quantum eraser'' with   time-binned photodetection, continuous photodetection should instead produce a joint probability given by (\ref{finprobi}) which displays steps in the coherent oscillations (see fig. (\ref{fig:proba})), 3) instead of  a ``quantum eraser'' with time resolution $\delta t \ll 1/\Delta \omega$ one could consider a ``quantum blurrer'' with a varying shutter time resolution. This serves as a window to let more which path information thereby suppressing the coherence in a controlled manner.

The experimental relevance of the questions studied in this article merit further study perhaps including alternative methods such as those of quantum open systems in terms of a master equation\cite{zoeller,dalibard} or ``quantum jumps'' followed by density matrix resetting as advocated in ref.\cite{heger}.

Entanglement and quantum correlations are becoming very important in many timely aspects of particle physics: in neutrino oscillations\cite{nu1,nu2} and in CP and T violation\cite{tviolent,Bo}. Recently the entanglement of neutral B-meson pairs produced from the (spontaneous) decay of a $\Upsilon(4S)$ resonance has been  exploited experimentally to unambiguously show time-reversal violation\cite{trevbo,babar} by tagging individual members of the correlated pairs. Therefore the interest on the dynamics of entanglement, the emergence of spontaneous coherence and quantum correlations is transcending disciplines and clearly merits deeper understanding.

\acknowledgements The author is deeply indebted to  A. Daley, G. Dutt, and D. Jasnow for their patience and enlightening comments and discussions and thanks J. Liang for an illuminating conversation and P. McMahon for bringing references\cite{qdotyama,qdotgao} to his attention. He acknowledges support from NSF-PHY-1202227.

\vspace{3mm}

\appendix
\section{Solutions for the coefficients in eqn. (\ref{psiAoft},\ref{psiDAoft})}\label{app}

\vspace{3mm}

The equations of motion for the coefficients in eqn. (\ref{psiAoft},\ref{psiDAoft})  are obtained from the Schroedinger equation in the interaction picture $d|\Psi(t)\rangle/dt = -iH_I(t)|\Psi(t)\rangle$   projecting on the corresponding states.

These simplify substantially from the following properties: $H_{AR}$ is the identity in the detector space $\{|e^d_j\rangle,|g^d\rangle \}$ and $H_{DR}$ is the identity in the NV-center basis $\{|A\rangle,|\pm 1\rangle \}$.

The equations of motion for the coefficients $C_{\vk,\pm}(t)$ feature contributions of the form
$$ \langle 1_{\vk,\pm};\mp 1;g^d|H_{DR}|\mp 1;e^d_j;0_\gamma\rangle \,D_{j,\mp}(t)$$
arising from the term $\sum_{j} \vec{d}^*_j\cdot \vec{E}^{(-)}(\vec{x}_d,t)|g^d\rangle\langle e^d_j|$ in $H_{DR}(t)$. Such term describes the de-excitation of the photodetector by spontaneous emission from an excited state $|e^d_j\rangle$ in which the NV-center states $|\pm 1\rangle $ are passive, this term is of higher order in dipolar couplings and under the assumption of very small detection efficiency as is the case experimentally (see below) it will be neglected\footnote{If necessary, this contribution can be obtained   from the unitarity condition $\langle \Psi(t)|\Psi(t)\rangle =1$.}, leading to the final form of the equations of motion
\bea  i\dot{C}_A(t) & = &   \langle A;0_\gamma|H_{AR}(t)|1_{\vk,+};-1\rangle \,{C} _{\vk,-}(t)+ \langle A;0_\gamma|H_{AR}(t)|1_{\vk,-};+1\rangle \,{C} _{\vk,+}(t) \label{dotcA2}\\ i\dot{C} _{\vk,+}(t) & = &
   \langle 1_{\vk,-};+1 |H_{AR}(t)|  A;0_\gamma\rangle \,{C} _{A}(t) \label{dotCp2}\\i\dot{C} _{\vk,-}(t) & = &
   \langle 1_{\vk,+};-1 |H_{AR}(t)|  A;0_\gamma\rangle \,{C} _{A}(t) \label{dotCm2}\\
 \dot{D}_{j,\pm}(t) & = & -i \frac{\vec{d}_j}{\sqrt{2}}\cdot  \langle 0_\gamma | \vec{E}^{(+)}_P(\vec{x}_d,t)|\sigma_\mp(t)\rangle \,\,e^{i\nu_j t} \label{dotDp}\,, \eea where the states $|\sigma_\mp(t)\rangle$ are given by (\ref{sigminplus}) with (\ref{Cpmfp}) evaluated at $\vec{x}_0=\vec{0}$. The solutions to eqns. (\ref{dotcA2},\ref{dotCp2},\ref{dotCm2}) are the same as (\ref{Caoftfp},\ref{Cpmfp}). Upon inserting these solutions   in the  matrix element  (\ref{dotDp}), we obtain in the Wigner-Weisskopf approximation
 \be -i \frac{\vec{d}_j}{\sqrt{2}}\cdot  \langle 0_\gamma | \vec{E}^{(+)}_P(\vec{x}_d,t)|\sigma_\mp(t)\rangle \,\,e^{i\nu_j t} = \kappa_j~ \sqrt{\frac{\Gamma_\pm}{2\pi}}~\delta^P_\mp \,e^{i\nu_j t_d}~e^{i(\nu_j-\Omega_\pm) (t-t_d)}\,e^{-\frac{\Gamma}{2}(t-t_d)} \,\Theta(t-t_d) ~~;~~t_d=\frac{x_d}{c}\label{mtxele}\ee where the constants $\kappa_j$ are proportional to $d_j/x_d$   with proportionality coefficients that result from angular and contour integration\footnote{For details see \cite{book1}.} and
 \be \delta^P_\mp = \Bigg\{  \begin{array}{c}
                               1 ~~\mathrm{for}~~P=H \\
                               \mp \, 1~~\mathrm{for} ~~P=V
                             \end{array} \,.\label{deltaP}\ee From this result we obtain

 \be D_{j,\pm}(t) =  i \kappa_j~ \sqrt{\frac{\Gamma_\pm}{2\pi}}~\delta^P_\mp \,e^{i\nu_j t_d}~ \frac{\Big[1-e^{i(\nu_j-\Omega_\pm+i\frac{\Gamma}{2}) (t-t_d)} \Big]}{\Big[\nu_j-\Omega_\pm + i\,\frac{\Gamma}{2} \Big]}~\Theta(t-t_d)\,.\label{djotasfin}\ee

\end{document}